\documentclass[10pt,twocolumn,letterpaper]{article}

\usepackage{cvpr}
\usepackage{times}
\usepackage{epsfig}
\usepackage{graphicx}
\usepackage{amsmath}
\usepackage{amssymb}
\usepackage{times}
\usepackage{epsfig}
\usepackage{kotex}
\usepackage{kotex-logo}
\usepackage[linesnumbered,ruled]{algorithm2e}
\usepackage{caption}
\usepackage{subcaption}
\usepackage{multirow}
\usepackage{url}
\usepackage{gensymb}
\newsavebox{\measurebox}
 \cvprfinalcopy

\usepackage[pagebackref=true,breaklinks=true,letterpaper=true,colorlinks,bookmarks=false]{hyperref}

\def\cvprPaperID{8767} % *** Enter the CVPR Paper ID here

\ifcvprfinal\fi
\begin{document}

%%%%%%%%% TITLE
\title{Multi-Temporal Recurrent Neural Networks For Progressive Non-Uniform\\
Single Image Deblurring With Incremental Temporal Training}

\author{Dongwon Park\thanks{equal contribution, $^{**}$ corresponding}\quad Dong Un Kang$^*$\quad Jisoo Kim\quad Se Young Chun$^{**}$\\
School of Electrical and Computer Engineering, UNIST, Republic of Korea\\
{\tt\small \{dong1,qkrtnskfk23,rlawltn1053,sychun\}@unist.ac.kr}
}

\maketitle
%\thispagestyle{empty}

%%%%%%%%% ABSTRACT
\begin{abstract}
Multi-scale (MS) approaches have been widely investigated 
for blind single image / video deblurring that sequentially recovers 
deblurred images in low spatial scale first and then in high spatial scale later with the output of lower scales.
MS approaches have been effective especially for severe blurs induced by large motions in high spatial scale
since those can be seen as small blurs in low spatial scale.
In this work, we investigate alternative approach to MS, called multi-temporal (MT) approach, for non-uniform single image deblurring.
We propose incremental temporal training with constructed MT level dataset from time-resolved dataset, 
develop novel MT-RNNs with recurrent feature maps, 
and investigate progressive single image deblurring over iterations.
Our proposed MT methods outperform state-of-the-art MS methods on the GoPro dataset in PSNR
with the smallest number of parameters.
\end{abstract}

%%%%%%%%% BODY TEXT
\section{Introduction}

\begin{figure}[!t]
	\centering
	\includegraphics[width=0.9\linewidth]{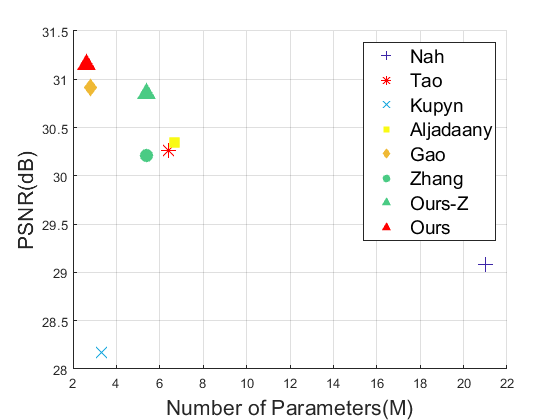}
			\vskip -0.1in
	\caption{Number of parameters (in Million) vs. PSNR (in dB) for different deblurring methods evaluated on the GoPro dataset. 
	Our proposed method (Ours) yielded the best PSNR (31.15dB) with the smallest number of parameters (2.6M) among all methods including
	Nah~\cite{nah2017deep}, Tao~\cite{tao2018scale}, Kupyn~\cite{kupyn2019deblurgan}, Aljadaany~\cite{aljadaany2019douglas}, Gao~\cite{gao2019dynamic} and Zhang~\cite{zhang2019deep}. `Ours-Z' is our MT approach with the network of Zhang~\cite{zhang2019deep}.}
	\label{fig:performance}
	\vskip -0.1in
\end{figure}

Blind single image deblurring is a challenging ill-posed inverse problem to recover the original sharp image from a given blurred image with or without estimating unknown non-uniform blur kernels and 
there has been much effort to tackle this problem. 
One is to simplify the given problem by assuming uniform blur and 
to recover both the latent ground truth image and the blur kernel~\cite{Fergus:2006ch,Shan:2008ef,Cho:2009dr,Xu:2010cr}. 
However, uniform blur is often not accurate enough to approximate the actual blur, and thus there has been much research on non-uniform blur by extending the degree of freedom of the blur model from uniform to non-uniform in a limited way  
compared to the dense matrix~\cite{Harmeling:2010we,Gupta:2010bv,Whyte:2010ct,Hirsch:2011fw,Xu:2013tl,Pan:2016ve}. 
Other non-uniform blur models have been investigated such as additional segmentations within which simple blur models were used~\cite{CouzinieDevy:2013bf,Kim:2013dg} or motion estimation based deblurs~\cite{Kim:2014gn,Kim:2015bw}.

\begin{figure*}[!t]
	\centering
	\includegraphics[width=0.95\linewidth]{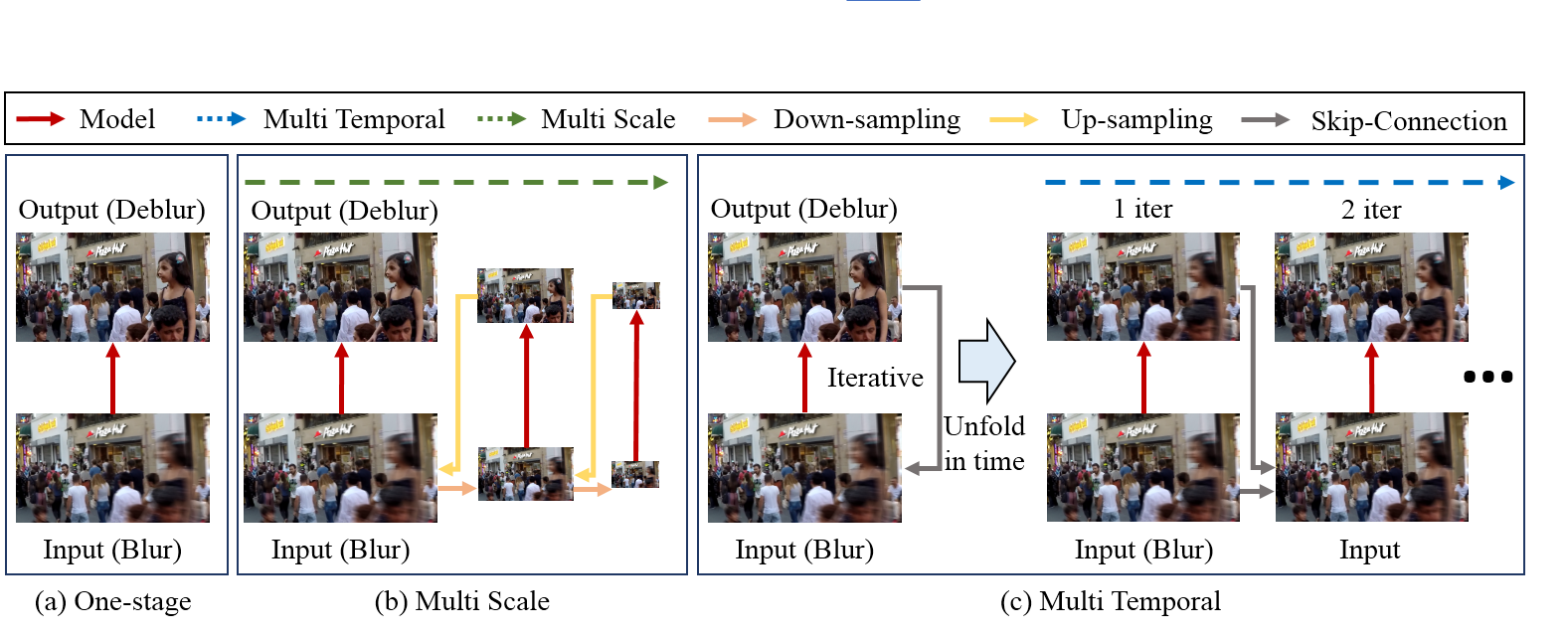}
			\vskip -0.1in
	\caption{Pipelines of three approaches for single image deblurring including single-scale (SS), multi-scale (MS) and our proposed multi-temporal (MT). MS and MT are progressively recovering images spatially and temporally, respectively.}
	\label{fig:pipelines}
	\vskip -0.1in
\end{figure*}

Recently, deep-learning-based approaches for single image deblurring have been proposed with excellent quantitative results and with fast computation time. There are largely two different ways of using deep neural networks (DNNs) for deblurring. One is to use DNNs to explicitly estimate non-uniform blurs~\cite{Sun:2015je,Chakrabarti:2016bj,Schuler:2016fk,Bahat:2017ez} and the other is to use DNNs to directly estimate the original sharp image without estimating blurs~\cite{Xu:2014wh,Kim:2017hk,Wieschollek:2017iy,Su:2017bk,nah2017deep,tao2018scale}. 
Most state-of-the-art methods such as~\cite{nah2017deep,tao2018scale} are estimating the original sharp image directly from the given blurred image (see Figure~\ref{fig:performance}).
Single-scale (SS) or one-stage approaches (Figure~\ref{fig:pipelines} (a)) are frequently used, but many state-of-the-art methods are using
multi-scale (MS) approaches (or coarse-to-fine) with down-scaled image(s) in spatial domain.

The MS approaches~\cite{nah2017deep,tao2018scale,gao2019dynamic} utilize down-scaled images to restore the latent sharp image progressively over scales as illustrated in Figure~\ref{fig:pipelines} (b).
%The down-scaled images are fed into the network for each stage. 
This approach makes use of the fact that blurs become relatively smaller as scale of image decreases~\cite{gao2019dynamic}. 
Thus, a DNN with MS approach is able to perform deblurring from large blur to small blur progressively. %It is efficient when handling severe blur. 
Recently, there have been some works on sharing network parameters of MS structures over scales~\cite{tao2018scale} or
efficiently sharing parameters except for feature extraction layers that are independent over scales
assuming that blurs are varied with scales~\cite{gao2019dynamic}. 
One drawback of typical MS approaches seems to lose much high-frequency information during down-sampling in a sub-optimal way for image deblurring considering the fact that strong edge information is important for reliable deblurring~\cite{Cho:2009dr,Xu:2010cr}. 

\begin{figure*}[!t]
	\centering
	\includegraphics[width=1.0\linewidth]{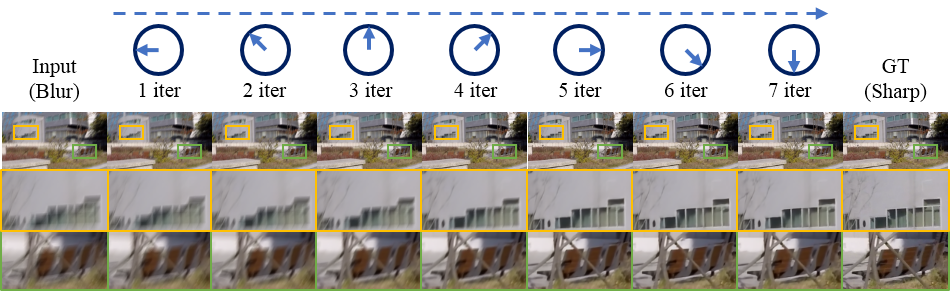}
			\vskip -0.1in
	\caption{Progressively deblurred images over iterations using our proposed MT-RNN with incremental temporal training.}
	\label{fig:iterative_deblurring_example}
	\vskip -0.1in
\end{figure*}

In this paper, we investigate an alternative approach, called multi-temporal (MT) approach, to MS approach for single image deblurring.
Instead of using down-scaled images, we exploit a typical dataset generation pipeline using high-speed camera to construct a blurred image
by averaging multiple frames of images such as the GoPro dataset~\cite{nah2017deep}.
We conjecture that recovering sharp latent images from mild blurs is easier than recovering them from severe blurs and propose
DNNs to deblur little by little as illustrated in Figure~\ref{fig:pipelines} (c).
Thus, our MT approach allows to use full image information in the original scale for reliable deblurring~\cite{Cho:2009dr,Xu:2010cr}
and to deblur for mild blurs at each iteration progressively (see Figure~\ref{fig:iterative_deblurring_example})
for potentially better performance than SS approach.
% Likewise, we narrow the temporal step to make the problem be easier, since the weak blur is close to latent image. 
Without any special parameter sharing schemes like~\cite{gao2019dynamic}, 
our proposed methods achieved state-of-the-art performance with the smallest number of parameters as shown in Figure~\ref{fig:performance}.  

Here is the summary of our contributions:
1) proposing MT approach with incremental temporal training for high-speed camera dataset to divide challenging severe blur into
a series of mild blurs and then deblur each mild blur progressively,
2) developing MT-recurrent neural network (RNN) with recurrent feature maps for blind single image deblurring, and
3) achieving state-of-the-art performance on GoPro dataset with the smallest number of parameters among recently proposed blind single image
deblurring methods.

\section{Related Works}

Conventional approaches to blind single image / video deblurring usually require to explicitly estimate blur kernels. There have been works on estimating uniform blurs using optimization with MS approach~\cite{Fergus:2006ch}, using a model of the spatial randomness of noise and a local smoothness prior~\cite{Shan:2008ef}, exploiting blurred strong edges to reliably estimate blur kernel~\cite{Cho:2009dr}, and developing a metric to measure the usefullness of image edges for blur kernel estimation~\cite{Xu:2010cr}.

There have also been many works on predicting non-uniform blurs assuming spatially linear blur~\cite{Harmeling:2010we}, simplified camera motion~\cite{Gupta:2010bv}, parametrized geometric model in terms of camera rotation velocity during exposure~\cite{Whyte:2010ct}, filter flow framework based blur model~\cite{Hirsch:2011fw}, $l_0$ sparsity for blurs~\cite{Xu:2013tl}, and dark channel prior~\cite{Pan:2016ve}. There was also an attempt to exploit multiple images from videos assuming spatially varying blur~\cite{Li:2010gy}. There have also been some works to utilize segmentation information by assuming uniform blur on each segmentation area~\cite{CouzinieDevy:2013bf} and to segment motion blur using optimization~\cite{Kim:2013dg}, to simplify motion model as local linear without segmentation using MS approach~\cite{Kim:2014gn}, and to use bidirectional optical flows for video deblurring~\cite{Kim:2015bw}.

Recently, many blind single image / video deblurring works employed DNNs for estimating blur kernels and/or original sharp images from given blurred input images. There are several works to predict non-uniform blur kernels explicitly: predicting the probabilistic distribution of motion blur at the patch level~\cite{Sun:2015je}, estimating the complex Fourier coefficients of a deconvolution filter~\cite{Chakrabarti:2016bj}, performing blur kernel estimation by division in Fourier space from extracted deep features~\cite{Schuler:2016fk}, and analyzing the spectral content of blurry image patches by reblurring them~\cite{Bahat:2017ez}. 

There are also many works to directly estimate the original sharp image from the given blurred input image without explicitly estimating non-uniform blur kernels. For video blind deblurring, there have been some works to exploit temporal information:
blending temporal information in spatio-temporal recurrent network for online video deblurring~\cite{Kim:2017hk}, taking temporal information into account with recurrent deblur network consisting of several deblur blocks~\cite{Wieschollek:2017iy}, and developing an encoder-decoder network with the input of multiple video frames to accumulate information across frames~\cite{Su:2017bk}.

There are a few works for blind single image deblurring without temporal information.
Xu \textit{et al.} proposed a direct estimation of the original sharp image based on optimization to approximate
deconvolution by a series of convolution steps using DNNs~\cite{Xu:2014wh}. Later, Nah~\textit{et al.} proposed a MS network architecture with Gaussian pyramid and MS loss functions~\cite{nah2017deep} and Tao~\textit{et al.} proposed convolution long short-term memory (LSTM)-based MS DNN for single image deblurring~\cite{tao2018scale}. Gao~\textit{et al.} proposed MS parameter sharing and nested skip connections~\cite{gao2019dynamic}. Zhang~\textit{et al.} proposed a deep multi-patch hierarchical network for different feature levels on the same resolution~\cite{zhang2019deep}. Aljadaany~\textit{et al.} proposed a learning both the image prior and data fidelity terms for single image deblurring~\cite{aljadaany2019douglas}. Kupyn~\textit{et al.}~\cite{kupyn2019deblurgan} proposes generative adversarial network (GAN) 
framework based on feature pyramid network (FPN) and relativistic discriminator~\cite{mao2017least} 
with a least-square loss~\cite{jolicoeur2018relativistic}.

Lastly, RNN plays an important role in using sequential data or iterative approach. 
Zhou~\cite{zhou2019spatio} proposed spatio-temporal variant RNN for video deblurring. 
RNN is often introduced to utilize previous frames effectively such as previous features using convolutional LSTM~\cite{tao2018scale}.
Similarly, we propose an approach that recurrently makes use of previous feature information for each iteration. 
However, unlike other RNN based approaches, our proposed methods use incremental temporal training procedure that does not train
from the most severe blur to the ground truth, but trains from more blured to less blurred image 
incrementally.

\section{Multi-Temporal (MT) Approach}

\begin{figure}[!b]
	\centering
		\vskip -0.1in
	\includegraphics[width=1.0\linewidth]{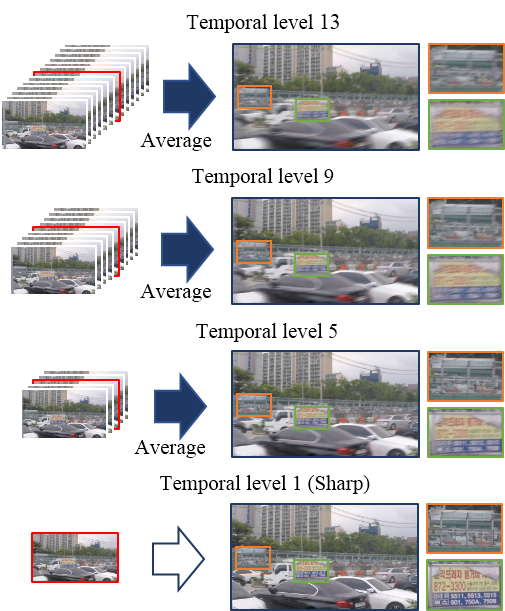}
			\vskip -0.1in
	\caption{Blurred images are generated by averaging multiple frames. More frames result in severe blurs. }
	\label{fig:data_generation}
\end{figure}

\subsection{GoPro Dataset}

The GoPro dataset consists of 15,000 sharp images (frames), captured by GoPro4 Hero Black camera (240 frame per sec), 
including 22 videos for training and 11 videos for testing~\cite{nah2017deep}. 
7-13 frames were averaged to yield blur-sharp image pairs 
where a middle image among multiple frames was selected as a ground truth as in Figure~\ref{fig:data_generation}.
%They~\cite{nah2017deep} took 240fps video and then averaged the successive latent image over varying number(7-13) for GoPro-Nah dataset. The middle of successive sharp image is corresponding to the latent sharp image on blurry image. 
Temporal level (TL) $N$ is defined to be a blurred image from $N$ frames. The GoPro dataset contains TL 7-13.

\subsection{Dataset For Incremental Temporal Training}
\label{subsec:gopro}

For the GoPro dataset with TL 1 (ground truth) and TL 7-13 pairs, 
we further generated data for MT approach and incremental temporal training.
For example, for a blurred image with TL 7, we generated intermediate blurred images with TL 1-13 as shown in Figure~\ref{fig:data_generation}.
Thus, our MT approach does not try to estimate TL 1 from TL 7 directly, but tries to estimate from TL 7 to TL 5, TL 5 to TL 3, and finally TL 3 to TL 1, progressively.

We quickly validated our conjecture for MT approach: will it be easier to estimate TL 1 from TL 7 than to estimate TL 1 from TL 5 or TL 3?
Table~\ref{table:IBL} shows the performance of U-Net~\cite{ronneberger2015u} that was trained only with one TL images for TL 3-13.
As TL increases, PSNR clearly decreases. Thus, our conjecture for MT approach seems reasonable.
%images(3-13). Each network is trained with only corresponding temporal level images. The structure of the network is architectureand identical over temporal levels. The PSNR is changed consistently with temporal level. It means that it is more difficult to perform deblurring at once, when blur becomes severe. Also, change of PSNR over blur level is linear. It shows that the network recognize certain temporal step. Therefore, handling severe blur step by step in several stage converges to better result.

\begin{table}[!h]
\centering
%		\vskip -0.1in
\caption{PSNR (dB) for single image deblurring using U-Net~\cite{ronneberger2015u} 
with input images with TL 3-13.}
		\vskip -0.1in
\begin{tabular}{|c|c|c|c|c|c|c|}
\hline
TL  & 3     & 5     & 7     & 9     & 11    & 13    \\ \hline
PSNR (dB) & 37.8 & 34.4 & 32.3 & 30.5 & 29.1 & 27.8 \\ \hline
\end{tabular}
\label{table:IBL}
		\vskip -0.1in
\end{table}

\subsection{Incremental Temporal Training}

Our training method is based on the dataset with more intermediate TL images as explained in Section~\ref{subsec:gopro}.
During training, our proposed network is recurrently iterated by 5 or 7. 
At iteration 1, we train the network with randomly selected temporal blurred images (TL 13 or 11 or 9 or 7) as inputs
and desired temporal blurred images (TL 11 or 9 or 7 or 5) as ground truth, respectively.
Note that the TL difference between input and ground truth is 2.
At the next iteration, the estimated image from iteration 1 is taken as input and desired temporal blurred image (TL 9 or 7 or 5 or 3) as ground truth. 
Similarly, other iterations are processed sequentially and take the estimated image from previous iteration as input and corresponding desired blurred or sharp image as ground truth. Finally, 1-3 more iterations of training to TL 1 as ground truth are repeated.
If the number of iterations is over 7, return to iteration 1 for training the model. 
Note that model parameters are shared and training is performed independently for each iteration.

\subsection{Progressive Deblurring With MT Approach}

The methods of Tao~\cite{tao2018scale} and Nah~\cite{nah2017deep} are based on MS approach for deblurring. 
The DNN of Tao~\cite{tao2018scale} shares parameters over scales that can be modeled as follows:
\begin{equation}
    \hat{I}^{j}, h^{j} = \mathrm{DNN}_\mathrm{Tao}(U(I^{0}),U(\hat{I}^{j-1}), U(h^{j-1}); \theta_\mathrm{Tao})
\end{equation}
Where $j$ refers to a scale where $j = 1$ represents the finest scale.
$I^0$ and $\hat{I}^j$ are blurred and estimated latent images at the $j$th scale, respectively.
$\mathrm{DNN}_\mathrm{Tao}$ is their MS network and $\theta_{Tao}$ is a set of parameters in their network. 
$h$ and $U$ are an intermediate feature map convolutional LSTM and a up-sampling operation by bilinear interpolation, respectively.

\begin{figure}[!t]
	\centering
	\includegraphics[width=0.95\linewidth]{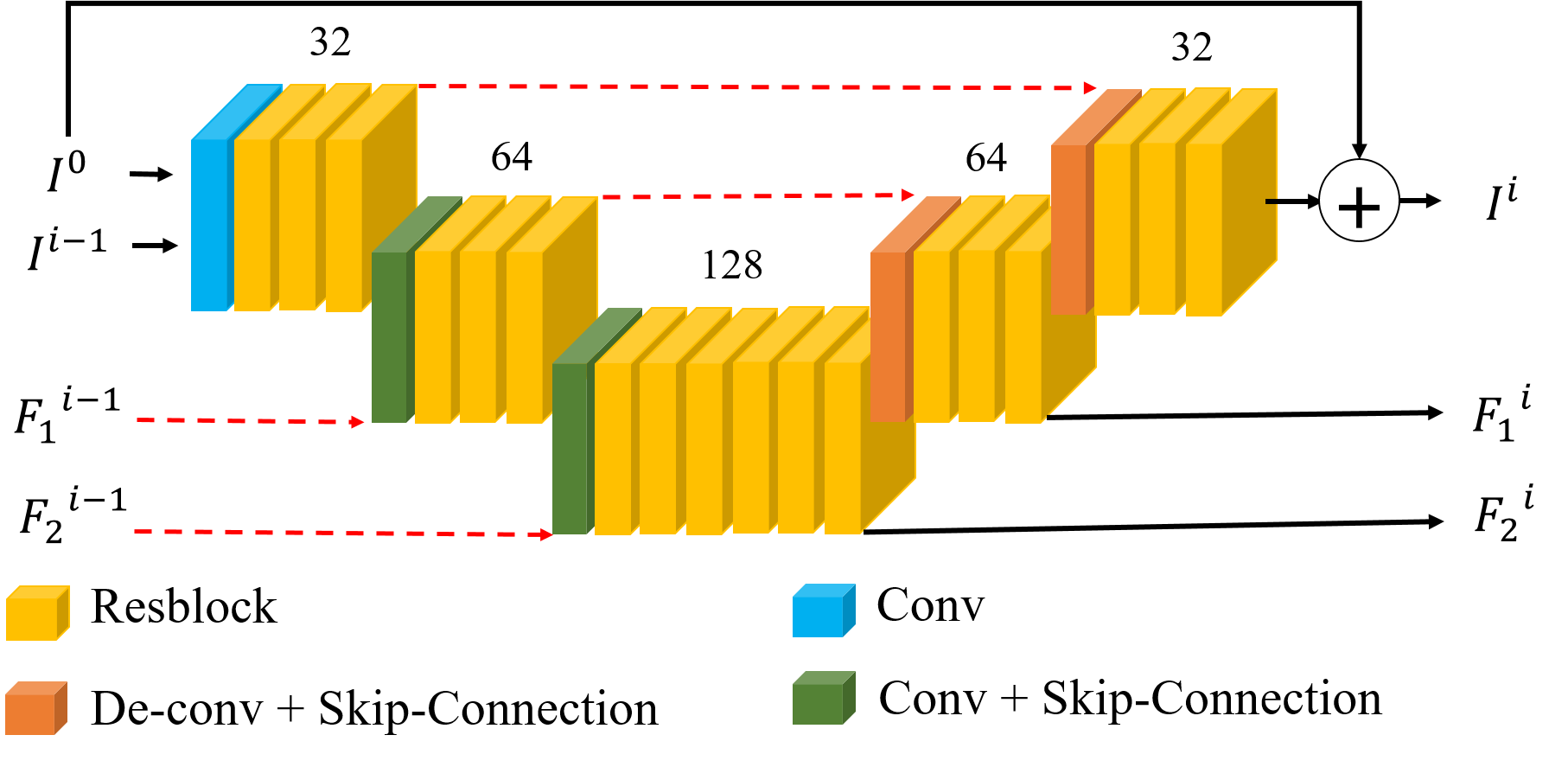}
			\vskip -0.1in
	\caption{Proposed architecture of MT-RNN.}
	\label{fig:model}
	\vskip -0.1in
\end{figure}

We propose MT-RNN with recurrent feature maps using temporal iterations that can be modeled as follows:
\begin{equation}
    \hat{I}^{i}, F^{i} = \mathrm{DNN}_\mathrm{Ours}(\hat{I}^{i-1}, I^{0}, F_{1}^{i-1}, F_{2}^{i-1}); \theta_\mathrm{Ours})
\end{equation}
where $i$ refers to an iteration number where $i=1$ represents the first iteration. 
$I^{0}$ is an input blurred image. $I^{i-1}$  and $I^i$ are blurred and estimated latent images at $i$th iteration, respectively. 
$F_{1}^{i-1}$ and $F_{2}^{i-1}$ are recurrent feature maps from $(i-1)$th decoder. %$F^{0}$ is the feature map of previous encoder. 
Since the network utilizes previous feature maps, the recurrent feature maps $F_{1}^{i-1}$ and $F_{2}^{i-1}$ move 
to the feature extraction layer in the next iteration. $\mathrm{DNN}_\mathrm{Ours}$ is our MT-RNN 
and $\theta_\mathrm{Ours}$ is a set of parameters in the network to be trained.
This model is illustrated in Figure~\ref{fig:model}.

\subsection{Proposed MT-RNN With Feature Maps}

Our proposed network is based on the network of Tao~\cite{tao2018scale}. 
Base model is U-Net architecture~\cite{ronneberger2015u} and consists of encoders and decoders as illustrated in Fig.~\ref{fig:model}. 
Each stage has 1 feature extraction layer and residual blocks (Resblocks) 
that is identical to the Resblock in~\cite{nah2017deep}
that consists of 32 channels, 64 channels and 128 channels at the top, middle and bottom encoder-decoders, respectively. %This can be modeled as follows.

\textbf{Residual learning} Kupyn~\cite{kupyn2019deblurgan} and Zhou~\cite{zhou2019spatio} utilize residual learning for deblurring. Both of them produce enhanced image and the network learns a residual image $I_{R}$ to correct the blurred image $I_{B}$
where $I_{deblur} = I_{B} + I_{R}$. Residual learning is efficient to train the network faster and resulting model generalizes better. In the deblurring problem, input and output are highly correlated. Therefore, the residual learning helps training the network.

We conducted an ablation study for residual learning.
In Figure~\ref{fig:model}, our proposed network takes $I^{0}$ and $\hat{I}^{i-1}$ as input and residual skip connection is linked with $I^{0}$. Two cases for $I_B$ was considered: $I^{0}$ and non residual skip. PSNR of connection with $I^{0}$ is higher than non residual learning by 0.15dB on the GoPro dataset with
intermediate TL images.

\textbf{Recurrent feature maps} As shown in Figure~\ref{fig:model}, recurrent features $F^{i-1}$ are from the last ResBlock of each decoder and are concatenated with the feature maps of previous encoder at feature extraction layer:
\begin{equation}
   F_{enc}^{i} = Cat(F^{i-1},f^{i})
\end{equation}
where $f^{i}$ is the feature map of previous encoder at the $i$th iteration. 
Estimated image $\hat{I}^{i-1}$ is concatenated with $I^{0}$:
\begin{equation}
   I_{cat}^{i} = Cat(\hat{I}^{i-1},I^{0})
\end{equation}
and then the encoder takes the $I_{cat}^{i}$ and $F_{encoder}^{i}$ as input.

Tao~\cite{tao2018scale} utilized convolutional LSTM for passing intermediate feature maps to the next spatial scale stage. 
Nah~\cite{Nah_2019_CVPR} also makes use of hidden state $h_{t-1}$ in RNN cell. 
Similarly, our network uses intermediate feature maps $F^{i-1}$ from decoder that
may include information about blur patterns and intermediate results for $I^{i}$. 
Thus, $F^{i-1}$ is utilized to encode $I^{i}$, having more details of blur patterns and other information for deburring. 
Using recurrent feature maps $F^{i-1}$ improves performance 
by 0.31dB.

\textbf{Loss Function}
We use $L1$ loss function that measures the difference between a restored image and its corresponding latent ground truth
normalized by channel, height and width of image. Ground truth images consists of TL 1-11 images.

\subsection{Convergence of MT-RNN over Iterations}

\begin{figure}[!t]
	\centering
	\includegraphics[width=0.8\linewidth]{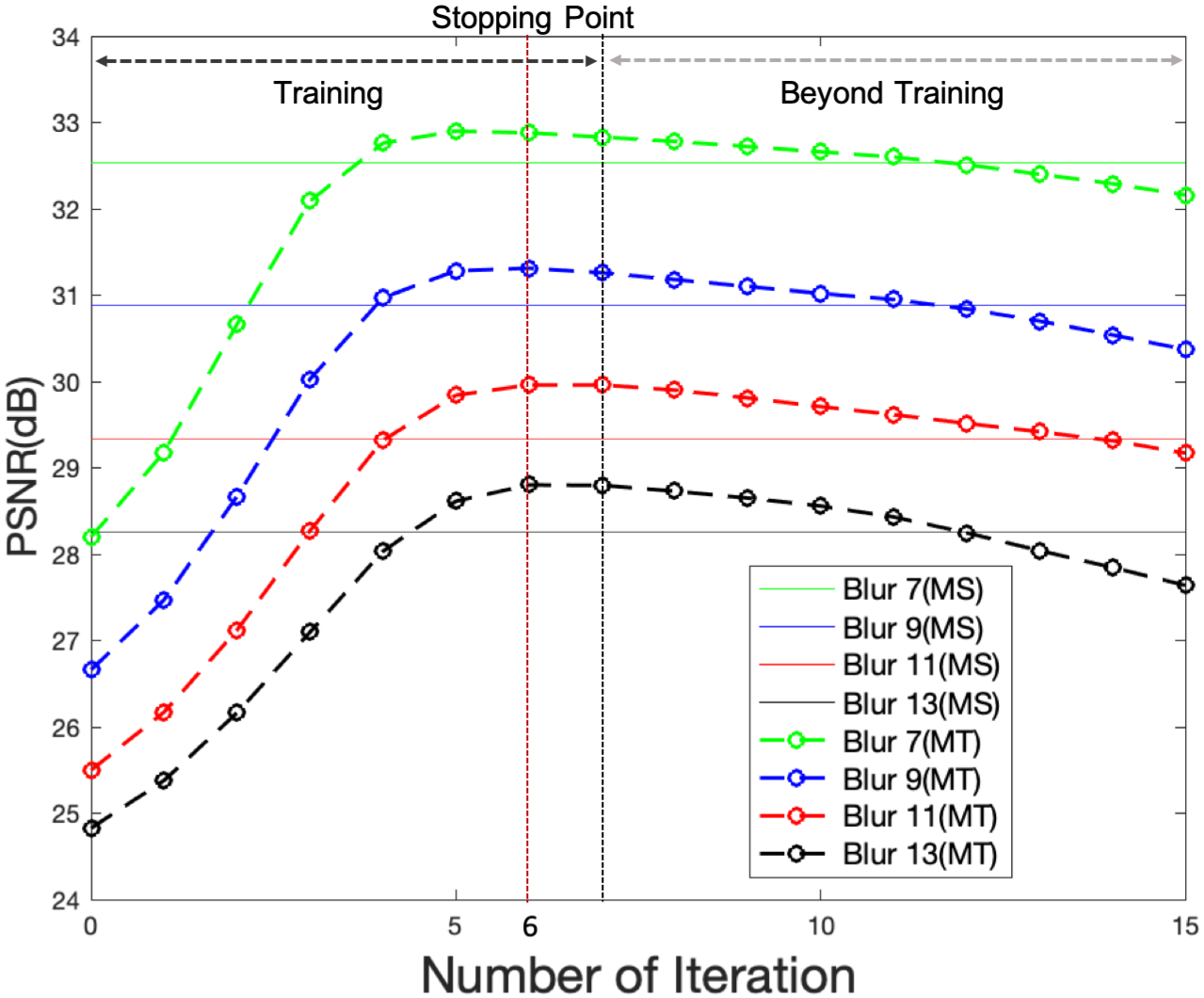}
			\vskip -0.1in
	\caption{Iteration vs. PSNR for our proposed MT-RNN trained using images with one of TL 7, 9, 11, 13.
	Corresponding models are trained only with each TL.}
	\label{fig:psnr}
	\vskip -0.1in
\end{figure}

Determining the number of iterations for MT-RNN is important for performance. We studied iteration vs. PSNR for the network that
was trained only with one type of TL images (e.g., TL 13) for all TL 7, 9, 11, 13.
Training was performed until the 7th iteration for all cases.
As illustrated in Figure~\ref{fig:psnr}, all networks yielded increased PSNR over iterations until 5th or 6th iterations,
and then decreased PSNR beyond training iterations.
From training procedure, iteration 6 was chosen and it was applied to all experiments for our methods.
Note that in all cases with different TL images, our proposed MT-RNN methods outperform state-of-the-art MS methods (Tao~\cite{tao2018scale}).

\section{Experiments}

\subsection{Dataset}

The GoPro dataset~\cite{nah2017deep} consists of 3214 blurred images with the size of 1280$\times$720 
that are divided into 2103 training images and 1111 test images.
In both validation and test sets, TL 7, 9, 11, 13 images were evenly distributed.
We generated more intermediate TL images along with the GoPro dataset so that
this new dataset consists of 5500 training, 110 validation and 1200 test images.

\subsection{Implementation Details}

We implement the proposed network on pytorch~\cite{paszke2017automatic}.
For fair comparisons, we evaluate our proposed method and state-of-the-art methods on the same machine with NVIDIA Titan V GPU. During training, Adam optimizer~\cite{kinga2015method} was used with learning rate $2\times10^{-4}$, $\beta_1=0.9$, $\beta_2=0.999$, and $\epsilon=10^{-8}$. For Tables~\ref{table:ablation}, \ref{table:blur_gap}, \ref{table:apply} and \ref{table:self_unsupervised}, 
total iteration was $92\times10^3$ with reducing learning rate by half every $46\times10^3$ iterations and the GoPro dataset with additional intermediate TL images was used.
% and we use the GoPro-Ours dataset. Also, The 
PSNR/SSIM were evaluated on python. Unlike others, for Table~\ref{table:benchmark}, total iteration is $46\times10^4$ 
with reducing learning rate by half every $46\times10^3$ iterations and the GoPro dataset was used.
% and GoPro-Nah is used only at evaluation of our proposed model. The 
PSNR/SSIM were evaluated on MATLAB. Patch size was 256$\times$256. 
Random crop, horizontal flip, and 90$^\circ$ rotation were used for data augmentation.
Note that since the number of channel is changed by concatenation to replace add operation of skip connection in~\cite{tao2018scale}, 1$\times$1 convolution was used.

\begin{table}[!b]
\centering
\caption{The ablation study with multi-scale approach (MS) and our proposed approach (MT). 
The components of ablation study are kernel size (K), residual learning (R), 
recurrent feature amp (F) and approaches (MS and MT). }
	\vskip -0.1in
\begin{tabular}{|l|c|c|c|c|c|c|c|c|}
\hline
\multicolumn{1}{|c|}{Approach}  & K & R  & F & PSNR & SSIM & Parm       \\ \hline
(a) MS~\cite{tao2018scale}     & 5      & X     & X             & 29.97 & 0.905 & 6.881      \\
(b) MS     & 3      & X   & X             & 30.10  & 0.906 & 2.584     \\
(c) MS     & 3      & O  & X             & 30.25  & 0.908 & 2.594     \\
(d) MT & 3      & O  & X             & 30.43  & 0.911 & 2.594\\
(e) \textbf{MT} & \textbf{3}      & \textbf{O}   & \textbf{O}             & \textbf{30.74}  & \textbf{0.917} & \textbf{2.635}\\
(f) MS+MT & 3      & O   & O             & 30.58  & 0.915 & 2.637\\ \hline
\end{tabular}
\label{table:ablation}
\end{table}

\subsection{Ablation Studies for Model Architecture}

We performed ablation studies from the base model (Tao~\cite{tao2018scale})
by adding our proposed components such as residual leaning and recurrent feature map.
Note that while Tao~\cite{tao2018scale} is spatially iterative due to MS approach, 
our proposed MT-RNN is temporally iterative due to our proposed MT structure.
Table~\ref{table:ablation} shows PSNR (dB), SSIM and number of parameter(Million) (denoted by Parm) 
along different components such as Approaches including MS approach or our proposed MT approach, 
%spatial iterative approach (denoted by M), temporal iterative approach (denoted by RMT), 
size of kernel (denoted by K), residual learning (denoted by R), recurrent feature map (denoted by F).
%on method, our proposed network is temporal iteration method. In order to confirm the contribution of each component, we perform ablation study: 
%Kernel size 5 vs 3; Non-Residual learning vs Residual learning; Multi Spatial iteration vs multi-temporal iteration; Lastly, Recurrent feature map. 
As a baseline MS network, Tao \cite{tao2018scale} was used as shown in Table~\ref{table:ablation} (a). 
Changing kernel size from 5 to 3 resulted in improved performance by 0.13dB,
and substantially decreased parameter size as in Table~\ref{table:ablation} (b).
Using residual learning instead of direct learning also improved performance as in Table~\ref{table:ablation} (c)
due to the effect of specified blurry region on back-propagation.
Our proposed MT approach improves performance over conventional MS approach using the same DNN as shown in Table~\ref{table:ablation} (d).
Further improvement was observed when using recurrent feature map as in Table~\ref{table:ablation} (e).
Lastly, %we finalized the experiment to see the effect of mixing up both MS and our MT approaches. It
it turns out that using MT alone is better than using both MS and MT in performance. Thus, 
temporal iterative approach helps the network to achieve high performance
as in Table~\ref{table:ablation} (f).

\subsection{Studies on Temporal Steps and Parameters}

We studied the effect of temporal steps on performance as shown in Table~\ref{table:blur_gap} (g), (h) and (e)
with one-stage SS approach (0 temporal step), 2 and 4 temporal steps, respectively.
Using MT approach yielded better performance than SS approach and 
using small temporal step was more advantageous than using larger temporal
step in performance even though computation time was increased.
Thus, we chose temporal step 2 as the best step size.

Then, we also investigated the effect of parameter size on performance.
Table~\ref{table:blur_gap} (e), (j), (k) show that the number of parameters is proportional to performance with the cost of increased computation.
While twice larger parameters in (k) did not improve performance much over (e), its computation time and memory were substantially increased.
Using half the parameter size in (j) did degrade performance substantially while computation speed of (j) is similar to (e).

\begin{table}[!b]
\centering
	\vskip -0.1in
\caption{Temporal steps (TS) and parameter sizes vs. performance. Parm in M, PSNR in dB, Time in sec.}
	\vskip -0.1in
\begin{tabular}{|c|c|c|l|l|l|}
\hline
TS & Iter&Parm & PSNR  & SSIM & Time \\ \hline
\multicolumn{1}{|l|}{(g) 0} & 1 & 2.63    & 29.93 & 0.904 & 0.005 \\
\multicolumn{1}{|l|}{(h) 4}    & 4 & 2.63    & 30.44 & 0.913 & 0.020\\
\multicolumn{1}{|l|}{(e) \textbf{2}}     & 6        & \textbf{2.63}    & \textbf{30.74} & \textbf{0.917} & \textbf{0.073}\\
\multicolumn{1}{|l|}{(j)  2}    & 6      & 1.46     & 30.21 & 0.908 & 0.060 \\
\multicolumn{1}{|l|}{(k) 2}    & 6      & 5.35     & 30.84 & 0.918 &0.290\\ \hline
\end{tabular}
\label{table:blur_gap}
\end{table}

\begin{table}[!t]
\centering
\caption{Applying our MT approach to other state-of-the-art methods, Kupyn~\cite{kupyn2019deblurgan} and Zhang~\cite{zhang2019deep}. 
Recurrent feature maps was not applied. PSNR in dB, Parm in M.}
	\vskip -0.1in
\begin{tabular}{|c|c|c|c|c|}
\hline
Method        & MT & PSNR & SSIM & Parm  \\ \hline
(l) Kupyn~\cite{kupyn2019deblurgan} & X         & 28.27  & 0.870 & 3.28\\
(m) \textbf{Kupyn}~\cite{kupyn2019deblurgan}& \textbf{O}         & \textbf{28.36}  & \textbf{0.872} & \textbf{3.28}\\
(n) Zhang~\cite{zhang2019deep}        & X         &  30.25 & 0.908 & 5.42\\
(o) \textbf{Zhang}~\cite{zhang2019deep}        & \textbf{O}         & \textbf{30.91}  & \textbf{0.918} & \textbf{5.43} \\ \hline
\end{tabular}
\label{table:apply}
\end{table}

\subsection{Our MT Approach to Other Deblur DNNs}

We investigate the feasibility of applying our proposed MT approach and incremental temporal training to other state-of-the-art deblurring
methods such as Kupyn~\cite{kupyn2018deblurgan} and Zhang~\cite{zhang2019deep} where the performances for them are
reported in Table~\ref{table:apply} (l) and (n), respectively.
Table~\ref{table:apply} (m) and (o) are performance results when using our proposed MT approach with incremental temporal training in
Kupyn and Zhang, respectively.
In both cases, our proposed approach successfully increased performance over the baselines.

\begin{table}[!b]
	\vskip -0.1in
\caption{Benchmarks on the GoPro dataset~\cite{nah2017deep} in terms of PSNR(dB), SSIM, parameter size (M) 
and run time (sec). `Ours-Z' indicates Zhang~\cite{zhang2019deep} with our MT approach.}
\label{table:GoPro}
	\vskip -0.1in
\centering
\begin{tabular}{|l|c|c|c|r|}
\hline
Method         & PSNR                  & SSIM                  & Parm(M)          &  Time           \\ \hline
Xu~\cite{Xu:2013tl}                       & 25.10                  & 0.890                 & -                     & 13.41s               \\
Kim~\cite{Kim:2014gn}                      & 23.64                 & 0.824                 & -                  & 1h                \\
Sun~\cite{Sun:2015je}                      & 24.64                 & 0.843                 & -                 & 20m                \\
Gong~\cite{gong2017motion}                     & 27.19                 & 0.908                 & -                   & -                     \\
Ram.~\cite{Ramakrishnan_2017_ICCV}                      & 28.94                 & 0.922                 & -                     & -                     \\
Nah~\cite{nah2017deep}                      & 29.08                 & 0.914                 & 21                        & 3.09s                \\
Kupyn~\cite{kupyn2018deblurgan}                    & 28.70                 & \textbf{0.958}                 & -                     & 0.85s                \\
Tao~\cite{tao2018scale}   & 30.26                 & 0.934                 & 6.4                      & 1.87s      \\ 
Kupyn~\cite{kupyn2019deblurgan}   & 28.17                 & 0.925                 & 3.3                & 0.04s      \\ 
Zhang~\cite{zhang2019deep}   & 30.21                 & 0.934                 & 5.4                      &  0.02s \\ 
Aljadaany~\cite{aljadaany2019douglas}   & 30.35                 & \textbf{0.961}                 &   6.7          & 1.2s      \\ 
Gao~\cite{gao2019dynamic}   & 30.92                 & 0.942                 & 2.8                       & 1.6s              \\ \hline
(o) Ours-Z                      & 30.78              & 0.940               & 5.4                & 2.08s              \\
(e) \textbf{Ours}                     & \textbf{31.15}              & {0.945}               & \textbf{2.6}                & \textbf{0.07s}              \\\hline
\end{tabular}
	\label{table:benchmark}
\end{table}

\begin{figure*}[!t]
	\centering
	\includegraphics[width=1.0\linewidth]{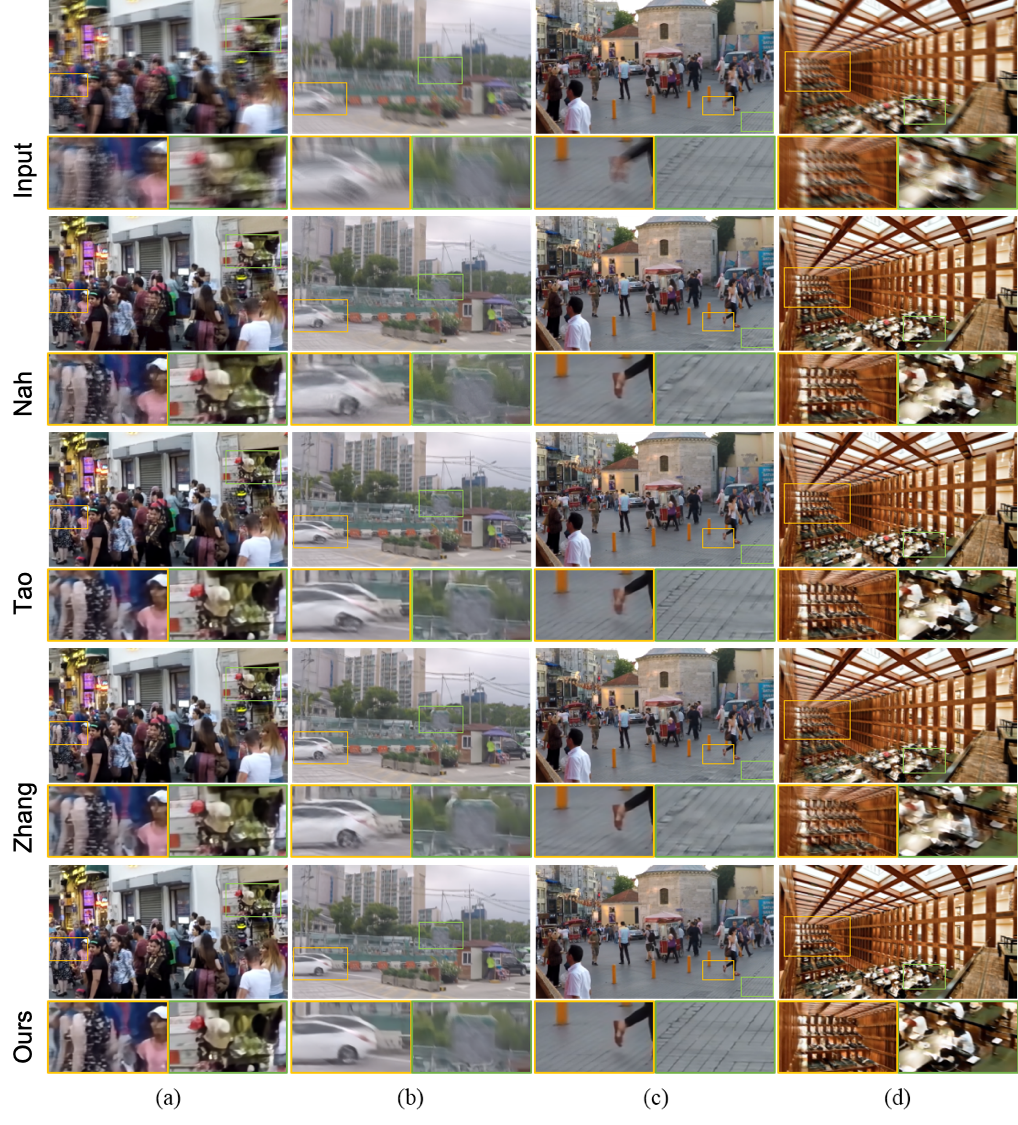}
%		\includegraphics[width=1.0\linewidth]{fig/reults_fig.png}
%			\vskip -0.1in
	\caption{Qualitative evaluations of various state-of-the-art methods as well as our proposed method on the GoPro dataset~\cite{nah2017deep}.
	Four input blurred images are on the 1st row, deblurred images of Nah~\cite{nah2017deep} on the 2nd row,
	deblurring results of Tao~\cite{tao2018scale} on the 3rd row, results of Zhang~\cite{zhang2018image} on the 4th row.
	Our results using MT-RNN are on the 5th row (bottom row). 
	Our proposed method yielded deblurred images that are visually better than the results of other state-of-the-art methods for
	all 4 image cases, especially for fine details of images.}
	\label{fig:benchmark}
%	\vskip -0.1in
\end{figure*}

\subsection{Benchmark Results}

We performed studies on the GoPro dataset~\cite{nah2017deep} for benchmarking. %~\cite{Su:2017bk}.
Tables~\ref{table:benchmark} presents quantitative results of our proposed methods and other state-of-the-art methods. 
Our proposed model, MT-RNN (e) 
were trained with the GoPro dataset along with intermediate TL images and achieve the best result (31.15 dB in PSNR) over other previous state-of-the-art methods on the GoPro test dataset (1111 images). 
Our MT approach for the network of Zhang~\cite{zhang2019deep} (o) also improved performance over the original network of Zhang.

Figure~\ref{fig:benchmark} shows qualitative evaluation in the case of four different models which are our proposed method (last row), the work of Nah~\cite{nah2017deep} (2nd row), the work of Tao~\cite{tao2018scale} (3rd row) and the work of Zhang~\cite{zhang2019deep} (4th row) for given blurred images (1st row). 
Qualitative results show that our proposed method outperforms other state-of-the-art methods visually.

\section{Discussion}

\begin{figure}[!t]
	\centering
	\includegraphics[width=1.0\linewidth]{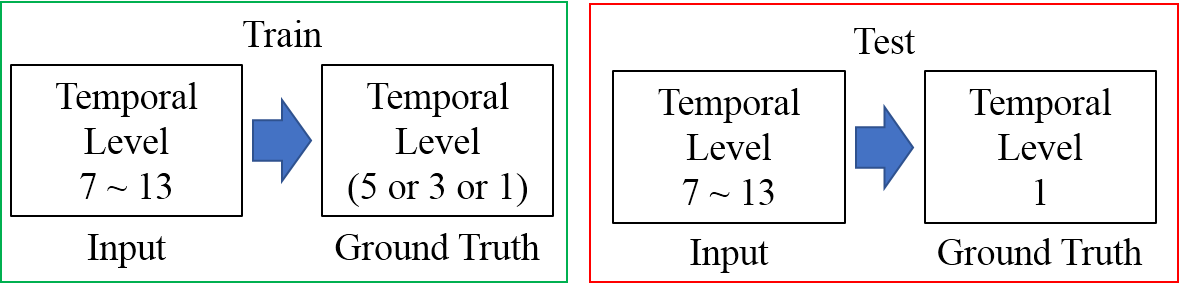}
			\vskip -0.1in
	\caption{Simulation setups for training with imperfect ground truth for deblurring and for testing with better ground truth than that during training.}
	\label{fig:self_unsupervised}
	\vskip -0.1in
\end{figure}

\textbf{On Imperfect Ground Truth}
Videos and images from high speed cameras often have mild blur assuming your subjects or objects move quickly.
Thus, obtaining perfect ground truth for single image deblurring problems is quite challenging.
Considering imperfect ground truth scenarios for single image deblurring, we perform experiments to observe the behaviors of our proposed
MT approaches and conventional MS approaches as illustrated in Figure~\ref{fig:self_unsupervised}.
During training, we used TL 1, 3, or 5 as ground truth and TL 7 - 13 as input images. % dataset.
Two approaches, MS and MT methods, were applied to these simulations: using TL 3 as ground truth and using TL 5 as ground truth for training.
\begin{table}[!b]
\centering \small
\caption{Performance results for different approaches (MT or MS) with different ground truth that contains certain TL images only (e.g., TL 3).
Our proposed MT approaches outperform MS approaches for all imperfect ground truth cases.
MT-n = MT approach with TL n data as ground truth.}
\begin{tabular}{|c|c|c|c|c|c|c|c|c|}
\hline
 & MT-5  & MT-3 & MT-1& MS-5  & MS-3 & MS-1 \\ \hline
PSNR & 29.47 & 30.28 & 30.74 & 28.61 & 29.76& 30.25 \\ \hline
SSIM & 0.895 & 0.901 & 0.917 & 0.876 & 0.897& 0.908 \\ \hline
\end{tabular}
	\label{table:self_unsupervised}
\end{table}

Table~\ref{table:self_unsupervised} shows that both MT and MS approaches yielded excellent performance assuming known perfect ground truth.
However, our proposed MT method yielded about 0.5dB better PSNR than the MS method.
When TL 3 images are given as ground truth, both MT and MS methods still yielded good performance. In this case, our MT approach yielded better
performance than conventional MS approach, that is also consistent with other performance comparison results in this paper.
One of the possible explanations on these results is that TL 3 images are already good enough as ground truth.
When TL 5 images are used as ground truth, the performance difference between MS and MT approaches became larger than other cases.
Thus, these preliminary results suggest that our proposed MT approaches may be more robust to imperfect ground truth dataset for deblurring than MS approaches.

\textbf{Decreasing PSNR Beyond Trained Iterations}
In Figure~\ref{fig:psnr}, MT-RNN yielded increasing PSNR during early iterations (usually, before 6 or 7 iterations) 
and then yielded decreasing PSNR later iterations.
To study the reasons for decreasing PSNR after stopping point, we visually investigated deblurred images from our proposed methods.
In images, we observed that there are often tiny artifacts appearing near the center of images. Then, as iteration increases, artifacts grows rapidly and they significantly

\textbf{Computation Time for ``Ours-Z''}
In the Table.~\ref{table:benchmark}, Ours-Z takes 2.08 seconds and iterates 6 times, while Zhang~\cite{zhang2019deep} takes 0.02 seconds without any iteration. Generally, running time of Ours-Z is expected around 0.02 seconds, but it takes 2.08 seconds in reality. For analyzing this issue, we measure the iteration time with one blurred image. The running time on a example image is 0.015 sec, 0.092 sec, 0.581 sec, 1.073 sec, 1.568 sec and 2.065 sec in the order of iterations.  Actually, the first iteration is similar with 0.02 seconds. However, after that, the running time increases exponentially.
Further investigation on this issue is necessary such as looking into GPU related issues. % and it may be caused by GPU problem. 

\textbf{Weight Sharing}
There are a few works on DNN based MS single image deblurring that share network weights across different scales in MS architecture~\cite{tao2018scale}
or that partially share network weights (except for feature extraction layers)~\cite{gao2019dynamic} 
so that the number of parameters is reduced significantly while performance is not degraded.
Note that our MT approach is similar to weight sharing across temporal iterations. 
However, partial shared parameters that may be much more efficient~\cite{gao2019dynamic} 
were not be investigated in MT structure. Thus, it will be interesting to further investigate partial weights schemes for MT approaches.

\section{Conclusion}

In this work, we investigate alternative approach to MS, called multi-temporal (MT) approach, for non-uniform single image deblurring.
We propose incremental temporal training with constructed MT level dataset from time-resolved dataset, 
develop novel MT-RNNs with recurrent feature maps, 
and investigate progressive single image deblurring over iterations.
Our proposed MT methods outperform state-of-the-art MS methods on the GoPro dataset in PSNR
with the smallest number of parameters.
%-------------------------------------------------------------------------
	\section*{Acknowledgments}
	
	This work was supported partly by 
	Basic Science Research Program through the National Research Foundation of Korea(NRF) 
	funded by the Ministry of Education(NRF-2017R1D1A1B05035810),
	the Technology Innovation Program or Industrial Strategic Technology Development Program 
	(10077533, Development of robotic manipulation algorithm for grasping/assembling 
	with the machine learning using visual and tactile sensing information) 
	funded by the Ministry of Trade, Industry \& Energy (MOTIE, Korea), and a grant of the Korea Health Technology R\&D Project 
	through the Korea Health Industry Development Institute (KHIDI), 
	funded by the Ministry of Health \& Welfare, Republic of Korea (grant number: HI18C0316).
%-------------------------------------------------------------------------

%-------------------------------------------------------------------------

{\small
\bibliographystyle{ieee_fullname}
\bibliography{egbib}
}

\end{document}